\documentclass[aps,prd,twocolumn,preprintnumbers,superscriptaddress,dblfloatfix,nofootinbib]{revtex4-1}
\usepackage[utf8]{inputenc}
\usepackage{lmodern}
\usepackage[dvipdfmx]{graphicx}
\usepackage{amsmath,amssymb,mhchem}
\usepackage{url}
\usepackage{color}
\usepackage{enumitem}
\usepackage{comment}
\usepackage{subfigure}
\usepackage[dvipsnames]{xcolor}
\usepackage[colorlinks=true,breaklinks=true]{hyperref}
\hypersetup{allcolors=[rgb]{0.0 0.0 0.6},linkcolor=[rgb]{0.75 0.05 0.05}}
\usepackage{bm}
\usepackage{epsfig}
\usepackage{amsmath}
\usepackage{amssymb}
\usepackage{slashed}
\usepackage{color}
\usepackage{accents}
\usepackage[dvipsnames]{xcolor}
\usepackage[colorlinks=true,breaklinks=true]{hyperref}
\hypersetup{allcolors=[rgb]{0.0 0.0 0.6},linkcolor=[rgb]{0.75 0.05 0.05}}
\usepackage{bm}
\usepackage{hyperref}
\usepackage{bashful}

\newcommand{\mnras}{Mon.~Not.~Roy.~Astron.~Soc.}
\newcommand{\physrep}{Phys.~Rep.}
\newcommand{\jcap}{JCAP}
\newcommand{\na}{New.~Astron.}
\newcommand{\pasj}{PASJ}
\newcommand{\apjl}{\apj~Lett.}

\newcommand{\hMpci}{$h\,$Mpc$^{-1}$}

\newcommand{\bx}{{\bf x}}

\newcommand{\bz}{{\bf z}}

\newcommand{\bk}{{\bf k}}

\newcommand{\dd}{\mathrm{d}}

\newcommand{\HL}[1]{{\textcolor{black}{#1}}}
\newcommand{\tnrv}[1]{{\textcolor{black}{#1}}}

\def\avrg#1{\left\langle #1 \right\rangle}

\begin{document}

\preprint{IPMU20-0075, YITP-20-85}

\title{Imprint of anisotropic primordial non-Gaussianity 
\\on halo intrinsic alignments
in simulations}

\author{Kazuyuki~Akitsu}
\email{kazuyuki.akitsu@ipmu.jp}
\affiliation{Kavli Institute for the Physics and Mathematics of the Universe (WPI), UTIAS \\The University of Tokyo, Kashiwa, Chiba 277-8583, Japan}

\author{Toshiki~Kurita}
\email{toshiki.kurita@ipmu.jp}
\affiliation{Department of Physics, The University of Tokyo, Bunkyo, Tokyo 113-0031, Japan}
\affiliation{Kavli Institute for the Physics and Mathematics of the Universe (WPI), UTIAS \\The University of Tokyo, Kashiwa, Chiba 277-8583, Japan}

\author{Takahiro~Nishimichi}
\affiliation{Center for Gravitational Physics, Yukawa Institute for Theoretical Physics, Kyoto University, Kyoto 606-8502, Japan}
\affiliation{Kavli Institute for the Physics and Mathematics of the Universe (WPI), UTIAS \\The University of Tokyo, Kashiwa, Chiba 277-8583, Japan}

\author{Masahiro~Takada}
\affiliation{Kavli Institute for the Physics and Mathematics of the Universe (WPI), UTIAS \\The University of Tokyo, Kashiwa, Chiba 277-8583, Japan}

\author{Satoshi~Tanaka}
\affiliation{Center for Gravitational Physics, Yukawa Institute for Theoretical Physics, Kyoto University, Kyoto 606-8502, Japan}

\date{\today}

\begin{abstract}
Using $N$-body simulations \HL{of cosmological large-scale structure formation,} for the first time, we
show that the anisotropic primordial non-Gaussianity (PNG) causes a scale-dependent modification, 
given by $1/k^2$ at small $k$ limit, in the three-dimensional power spectra of halo shapes (intrinsic alignments), whilst 
the conventional power spectrum of halo number density field remains unaffected. We discuss that wide-area imaging and spectrocopic surveys observing the same region of the sky allow us to 
constrain the
quadrupole PNG coefficient $f_{\rm NL}^{s=2}$ at 
a precision comparable with 
or better than 
that of the cosmic microwave background.
\end{abstract}

\maketitle

\section{Introduction}
An observational exploration of non-Gaussianity in the primordial perturbations, which are the seeds of cosmic structures, gives a powerful test of the physics in the early universe such as inflation \citep{Maldacena:2003,PNG_review,Arkani-Hamed_Maldacena:2015}. 
The cosmic microwave background (CMB) anisotropies and wide-area galaxy surveys can be used to pursue the primordial non-Gaussianity (PNG) from their observables \citep{Komatsu_Spergel:2001,Planck_PNG:2019,Dalal_etal:2008,Slosar_etal:2008,Shiraishi_etal:2013}
and these two carry complementary information.

Suppose that $\Phi(\bx)$ is the primordial potential field. 
The simplest PNG model is a local-type one, and its bispectrum is generally, as in given by Refs.~\cite{Shiraishi_etal:2013,Schmidt_etal:2015}:
\begin{align}
&B_\Phi(\bk_1,\bk_2,\bk_3)\nonumber\\
&\hspace{0.5em}=2\!\!\sum_{\ell=0,1,2,\cdots}f_{\rm NL}^{s=\ell} 
\left[
{\cal L}_\ell(\hat{\bk}_1\cdot\hat{\bk}_2)
P_\phi(k_1)P_\phi(k_2)+\mbox{2 perms.}
\right], 
\label{eq:bispectrum_s}
\end{align}
where $\hat{\bk}\equiv \bk/k$, $P_\phi(k)$ is the power spectrum of a Gaussian field, denoted as $\phi(\bx)$, and ${\cal L}_\ell$ is the Legendre polynomial 
of order $\ell$; ${\cal L}_0(\mu)=1$ and ${\cal L}_2(\mu)=(3\mu^2-1)/2$. The coefficient,  $f^{s=\ell}_{\rm NL}$, is a parameter to characterize 
the amplitude of the local PNG at each order $\ell$. 
Due to the orthogonality of the Legendre polynomials ${\cal L}_\ell$, the PNG modes of different $\ell$ are independent with each other, and are expected to carry complementary information on the physics in the early universe, if detected or constrained separately. 
The {\it isotropic} PNG model with $s=0$ has been well studied in the literature \citep{PNG_review,Dalal_etal:2008}.
The reality condition of $\phi(\bx)$ ensures that 
the odd multipoles should vanish 
in the squeezed limit, where one of wavevectors 
is much smaller than the other two.
Thus, in this
paper we focus on the \textit{anisotropic} PNG described by the $s=2$ term in the above bispectrum,
which is the leading-order anisotropic PNG model among PNGs
that have greater amplitudes in the squeezed limit\footnote{\HL{Our notation 
$f_{\rm NL}^{s=2}$ is different from the notation $A_2$ used in 
Ref.~\cite{Schmidt_etal:2015}; the relation is $A_2 = 4f_{\rm NL}^{s=2}$.}}.

The anisotropic PNG can be generated in
several inflationary scenarios:
the solid inflation~\cite{Endlich_etal:2012}, the non-Bunch-Davies initial states~\cite{Agullo_Shandera:2012},
and the existence of vector fields~\cite{Bartolo_etal:2012,Shiraishi:2012,Shiraishi_etal:2012,Shiraishi_etal:2013,Bartolo_etal:2015} and
higher-spin fields~\cite{Arkani-Hamed_Maldacena:2015,Lee_etal:2016,Franciolini_etal:2018} in the inflationary epoch.
Although 
the predicted bispectrum 
generally has a particular scale dependence such as
${\cal L}_\ell(\hat{\bk}_1\cdot\hat{\bk}_2)
\to\left(k_1/k_2\right)^{\Delta_{\ell}}{\cal L}_\ell(\hat{\bk}_1\cdot\hat{\bk}_2)$
in Eq.~\eqref{eq:bispectrum_s},
we consider a model with 
$\Delta_2=0$ 
for simplicity in this paper.

\HL{Hence the purpose of this paper is to study how the anisotropic PNG (the term 
$f^{s=2}_{\rm NL}$ in Eq.~\ref{eq:bispectrum_s}) affects the power spectrum of galaxy shapes, the so-called intrinsic alignment (IA), that is measured from wide-area galaxy surveys \cite{Hirata_Seljak:2004,Schmidt_etal:2015,Kurita_etal:2020}. To do this, we for the first time run $N$-body
simulations adopting the anisotropic PNG initial conditions, and then measure the three-dimensional power spectrum 
of halo shapes, as a proxy of the IA observables of galaxy shapes. For
completeness
of our discussion we also run $N$-body simulations using the Gaussian and isotropic PNG ($f^{s=0}_{\rm NL}$) initial conditions, and then compare the results for the IA power spectra and the power spectrum 
of halo number density field. Then we perform the Fisher forecast to estimate an ability of wide-area galaxy survey for constraining the anisotropic PNG.}

The structure of this paper is as follows. In Section~\ref{sec:preliminaries} we briefly review 
the PNG initial conditions, the IA effect and the expected effect of anisotropic PNG on the IA. 
In Section~\ref{sec:simulations} we describe details of $N$-body simulations with Gaussian 
and PNG initial conditions we use in this paper. In Section~\ref{sec:results} we show the main results of this paper, i.e. the IA power spectrum measured from the anisotropic PNG simulations, and compare the results with the power spectra for the Gaussian and isotropic PNG simulations. 
Section~\ref{sec:discussion} is devoted to discussion.
Throughout this paper, unless otherwise stated, we employ the flat-geometry $\Lambda$CDM cosmology 
\tnrv{as a}
fiducial model, which is consistent with the {\it Planck} CMB data \citep{Planck2015_cosmo}.
The model is characterized by $\Omega_{\rm m}=0.3156$ for the matter density parameter,
 $\omega_{\rm b}(\equiv \Omega_{\rm b}h^2)=0.02225$, $\omega_{\rm c}(\equiv \Omega_{\rm c}h^2)=0.1198$ for the physical density parameters of baryon and CDM,  
and $n_{\rm s}=0.9645$ and $\ln(10^{10}A_{\rm s})=3.094$ for the tilt and amplitude parameters of the primordial curvature power spectrum. This fiducial model is the same as that used in \citet{Nishimichi_etal:2019}.

\section{Preliminaries}
\label{sec:preliminaries}

\subsection{Nonlinear transformation from anisotropic PNG}
To 
generate numerical realizations of random fields with the PNG given by the $s=2$ term in the bispectrum (Eq.~\ref{eq:bispectrum_s}),
we consider the following nonlinear transformation of $\phi$:
\begin{align}
\Phi(\bx)=\phi(\bx)+\frac{2}{3}f_{\rm NL}^{s=2}
\sum_{ij}
\left[(\psi_{ij})^2(\bx)-\langle (\psi_{ij})^2
\rangle\right],
\label{eq:Phi_def}
\end{align}
where $\psi_{ij}$ is a
trace-less tensor that has the same dimension as $\phi$,
defined as
\begin{align}
\psi_{ij}&\equiv \frac{3}{2}\left[\frac{\partial_i\partial_j}{\partial^2} -\frac{1}{3}\delta_{ij}^{\rm K}\right]\phi\nonumber\\
&=\int\!\frac{\mathrm{d}^3\bk}{(2\pi)^3} \frac{3}{2} \left(\hat{k}_i\hat{k}_j-\frac{1}{3}\delta^{\rm K}_{ij}\right)\phi(\bk)e^{i\bk\cdot\bx},
\label{eq:psi_def}
\end{align}
where $\delta^{\rm K}_{ij}$ is the Kronecker delta function. 
One can easily confirm that the non-Gaussian field $\Phi$ leads to the bispectrum with $s=2$
in Eq.~(\ref{eq:bispectrum_s}).

For galaxy surveys, the mass density fluctuation field, $\delta(\bx)$, instead of the primordial potential $\Phi(\bx)$, is more relevant 
for observables. These fields in the linear regime are related to each other via 
$\delta(\bk)={\cal M}(k,z)\Phi(\bk)$, where ${\cal M}(k,z)\equiv (2/3)k^2T(k)D(z)/(\Omega_{\rm m0}H_0^2)$, 
with $T(k)$ and $D(z)$ denoting the transfer function and
the linear growth
factor, respectively.
As discussed in Ref.~\cite{Schmidt_etal:2015}, 
in the presence of the above PNG,
the amplitude of the local small-scale power spectrum at the position $\bx$ has a modulation depending on
the long-wavelength potential $\psi^{L}_{ij}$
as 
\begin{align}
P_{\delta}(\bk|\bx)|_{\psi^L_{ij}} = \left[1+4f_{\rm NL}^{s=2}\psi^L_{ij}(\bx)\hat{k}^i\hat{k}^j \right]P_\delta(k),
\label{eq:P_phi_picture}
\end{align}
where $\bk$ is
a short-wavelength mode. Since $\psi_{ij}^L$ is 
a trace-less tensor, it
causes a
quadrupolar modulation in the power of 
short mode fluctuations.

\subsection{Intrinsic alignment and PNG}
The linear intrinsic alignment (IA) model \cite{Catelan_etal:2001,Hirata_Seljak:2004,Schmidt_etal:2015}
predicts that
the shapes of galaxies originate 
from the gravitational tidal field as
\begin{align}
\gamma_{ij}(\bx)=b_K K_{ij}(\bx),
\label{eq:gammaij_Gaussian}
\end{align}
where $\gamma_{ij}$ is the ($3\times 3$)-tensor characterizing
the shape of each galaxy 
and $K_{ij}$ is the 
tidal field at the galaxy's position. 
We define the tidal field as 
$K_{ij}=(\partial_i\partial_j/\partial^2 - \delta^{\rm K}_{ij}/3)\delta$ so that $K_{ij}$ has the same dimension as that of the mass density fluctuations (so $K_{ij}$ is a dimension-less quantity).
This relation holds on
scales 
sufficiently larger than the reach of galaxy and halo formation physics. 
Here $b_K$ is the linear shape ``bias'' coefficient, which can be interpreted as
a response of the galaxy
shape to the long-wavelength 
tidal field,
whereas the linear ``density'' bias parameter $b_1$ gives a 
response of the galaxy number density to the long-wavelength 
mass density fluctuation 
\cite{Kaiser:1984,Mo_White:1996,bias_review}
\footnote{\HL{The linear shape bias $b_K$ is related to the conventionally used linear alignment coefficient $C_1$, used in the literature \citep[e.g.][]{Kurita_etal:2020},  
through $b_K = - a^3 C_1 \bar{\rho}(a)/D(a)$.}}.
For adiabatic, Gaussian initial conditions, $b_K$ takes a constant (scale-independent) value at the limit of a sufficiently large smoothing scale or $k\rightarrow 0$ 
in Fourier space, and the value varies with the type of galaxies. However, the \textit{anisotropic} PNG breaks the condition, and causes a characteristic 
scale-dependent modification in $b_K$, 
as the \textit{isotropic} PNG does for the density tracers \citep{Dalal_etal:2008}.

As we discussed in Eq.~(\ref{eq:P_phi_picture}), the anisotropic PNG 
induces the coupling between the local tidal field, $K_{ij}$, and the long-wavelength quadrupole potential, $\psi_{ij}$.
Similarly to the effect of isotropic PNG on the density distribution
of galaxies,
this mode-coupling leads to a scale-dependent modification in the IA
of galaxy shapes as pointed out by Ref.~\cite{Schmidt_etal:2015}:
\begin{align}
\gamma_{ij}(\bk)\simeq \left[b_K+12 b_\psi f_{\rm NL}^{s=2}{\cal M}^{-1}(k)\right]K_{ij}(\bk), 
\label{eq:gammaij_PNG_model}
\end{align}
where $b_\psi$ is a parameter to characterize the response of galaxy shapes to the long-wavelength quadrupole potential, defined 
as $b_\psi \equiv \partial \gamma_{ij}/\partial (2 f_{\rm NL}^{s=2}\psi_{ij})$. 
The second term on the r.h.s. shows that 
 the anisotropic PNG induces a scale-dependence of $1/k^2$ in the 
IA effect at very small $k$, as in the effect of the local-type isotropic PNG on the galaxy density bias parameter \cite{Dalal_etal:2008}.
In the following
we treat $b_K$ and $b_\psi $ as  free parameters, and then estimate their values (the value of $b_\psi$ for the first time) 
for a sample of halos from $N$-body simulations adopting the Gaussian and the anisotropic PNG initial conditions, respectively. 
If we use the peak theory for the nearly random, Gaussian field, extending the formula in Refs.~\cite{BBKS:1986,Bond_Efstathiou:1987}, 
we might be able to estimate a relation between $b_K$ and $b_\psi$ for halos. However, this is beyond the scope of this paper, and will be our future work.
We also note that an apparent infrared divergence at the limit $k\rightarrow 0$ should be restored if properly taking into account the finite survey region and 
relativistic effects \citep[e.g. see Refs.][for the discussion on the density bias parameter]{Wands_Slosar:2009,Donghui_etal:2012,Baldauf_etal:2011}. Since we are interested in the IA effect on subhorizon scales, we can safely ignore the relativistic effects. 

\section{Numerical implementations}
\label{sec:simulations}

\subsection{Initial conditions and simulations}

To generate the initial conditions for $N$-body simulations with the anisotropic PNG,
we modified \texttt{2LPTic}, developed in Ref.~\cite{Crocce_etal:2006,Scoccimarro_etal:2012}.
First, in Fourier space we generate
a Gaussian random field $\phi(\bk)$ using the assumed $P_\phi$,
and prepare the auxiliary field $\psi_{ij}(\bk)$ 
according to Eq.~(\ref{eq:psi_def}).
Then Fourier transforming $\phi(\bk)$ and $\psi_{ij}(\bk)$ to real space, 
we construct the non-Gaussian field $\Phi({\bf x})$ following Eq.~(\ref{eq:Phi_def}). We solve the Lagrangian dynamics up to the second order based on the non-Gaussian field $\Phi$ and the matter transfer function computed by \texttt{CLASS} \cite{Blas_etal:2011}. 
Throughout this paper we employ
a flat $\Lambda$CDM cosmology with the parameters summarized at the bottom of the introduction that is consistent with the {\it Planck} data
\cite{Planck2015_cosmo}.
We confirmed explicitly that the bispectrum measured from the
$\Phi$ field generated with this procedure is consistent with 
the $s=2$ term of Eq.~(\ref{eq:bispectrum_s}). 

We then evolve the particle distribution using a newly developed $N$-body solver based on the Tree Particle-Mesh (PM) scheme \cite{Nishimichi_etal:2020}. It is based on a general-purpose framework for particle methods, \texttt{FDPS}~\cite{Masaki_etal:2016,Namekata_etal:2018}, with the PM part originally implemented in \texttt{GreeM}
\cite{Yoshikawa_Fukushige:2005,Ishiyama_etal:2009,Ishiyama_etal:2012}. We further accelerate the calculation of gravitational force term with a 512-bit SIMD instruction set based on Intel AVX-512 
in a similar manner as in the \texttt{Phantom-GRAPE} library \cite{Tanikawa_etal:2012,Tanikawa_etal:2013,Yoshikawa_Tanikawa:2018} and optimize the memory footprint for efficient execution in high-performance parallel environments. The final accuracy of this code is tuned such that it reproduces the matter power spectrum from a \texttt{Gadget2}~\cite{gadget2} run started from an identical initial condition with the accuracy parameters used in \cite{Nishimichi_etal:2019}, to within one percent up to the particle Nyquist frequency. 
We adopt $N_{\rm part}=2048^3$ particles and $4.096~h^{-1}{\rm Gpc}$ and $2.048~h^{-1}{\rm Gpc}$ for the comoving simulation box size. The particle masses are $m_{\rm p}\simeq 7.0\times 10^{11}~h^{-1}M_\odot$ and $m_{\rm p}\simeq 8.8\times 10^{10}~h^{-1}M_\odot$
for $4.096~h^{-1}{\rm Gpc}$ and $2.048~h^{-1}{\rm Gpc}$ boxes respectively.
For comparison, we also run
simulations for
a Gaussian initial condition
and the isotropic ($s=0$) PNG model, using the same initial seeds. 

In summary we run 
11 simulations in total. We run 6 large-box simulations, which contain 
one Gaussian simulation and
5 PNG simulations with  
$f^{s=0}_{\rm NL}=500$ and $f_{\rm NL}^{s=2}=\pm100$ and $\pm 500$,
 for $4.096~h^{-1}{\rm Gpc}$ box. 
\HL{In addition, to study the effect of simulation resolution, 
 we also use 5 small-box simulations, which contain
one Gaussian simulation and
4 PNG simulations with  
$f_{\rm NL}^{s=2}=\pm100$ and $\pm 500$, for $2.048~h^{-1}{\rm Gpc}$ box.}
We use
the shapes of halos identified by {\tt Rockstar} \citep{Behroozi_etal:2013}, as a proxy of the galaxy IA effect, and also 
use the {\tt Rockstar} output to infer the 
virial mass of each halo, denoted as $M_{\rm vir}$ \citep{Kurita_etal:2020}.

\begin{figure*}
    \begin{center}
        \includegraphics[width=2.0\columnwidth]{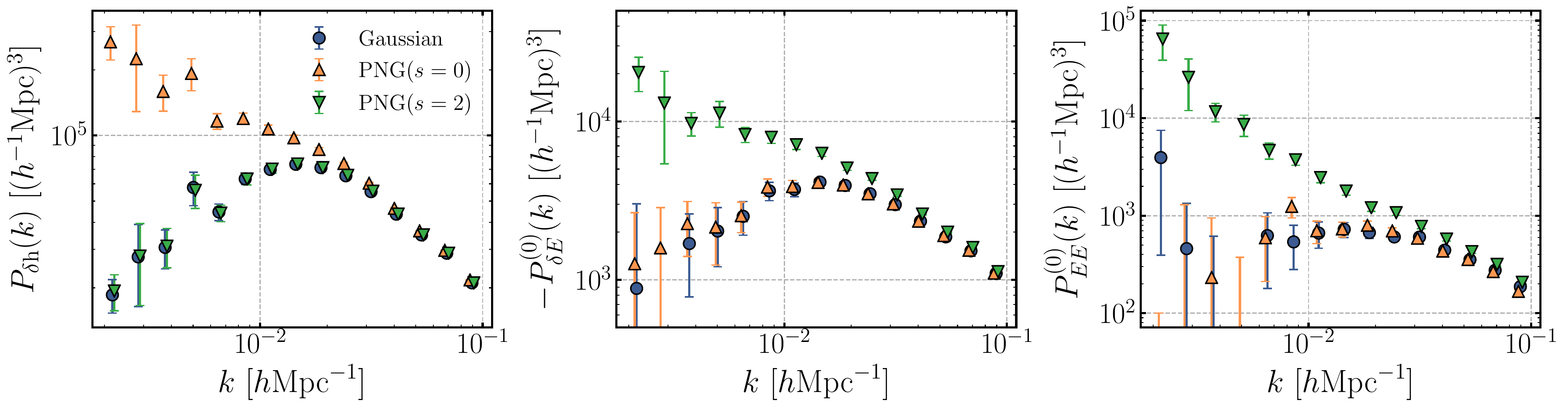}
        \caption{The 
        matter-halo power spectrum (left panel), the monopole moment of the cross-power spectrum of matter and halo shapes (middle),
        and the monopole moment of shape-shape auto-spectrum (right)
         for various initial conditions; Gaussian (blue), isotropic PNG (orange) and anisotropic PNG (green) initial conditions, respectively.
        Here we assume ($f_{\rm NL}^{s=0}, f_{\rm NL}^{s=2})=(500,0)$ or $(0,500)$  for the isotropic or anisotropic PNG case (Eq.~\ref{eq:bispectrum_s}), respectively.  
        These are measured for the halo sample with $M_{\rm vir} > 10^{14} h^{-1}M_\odot$ at $z=0$. The errorbars denote the Gaussian errors for a volume of 
        $V=69~(h^{-1}{\rm Gpc})^3$.
        }
        \label{fig:ps}
    \end{center}
\end{figure*}
\begin{figure}
    \begin{center}
        \includegraphics[width=1.0\columnwidth]{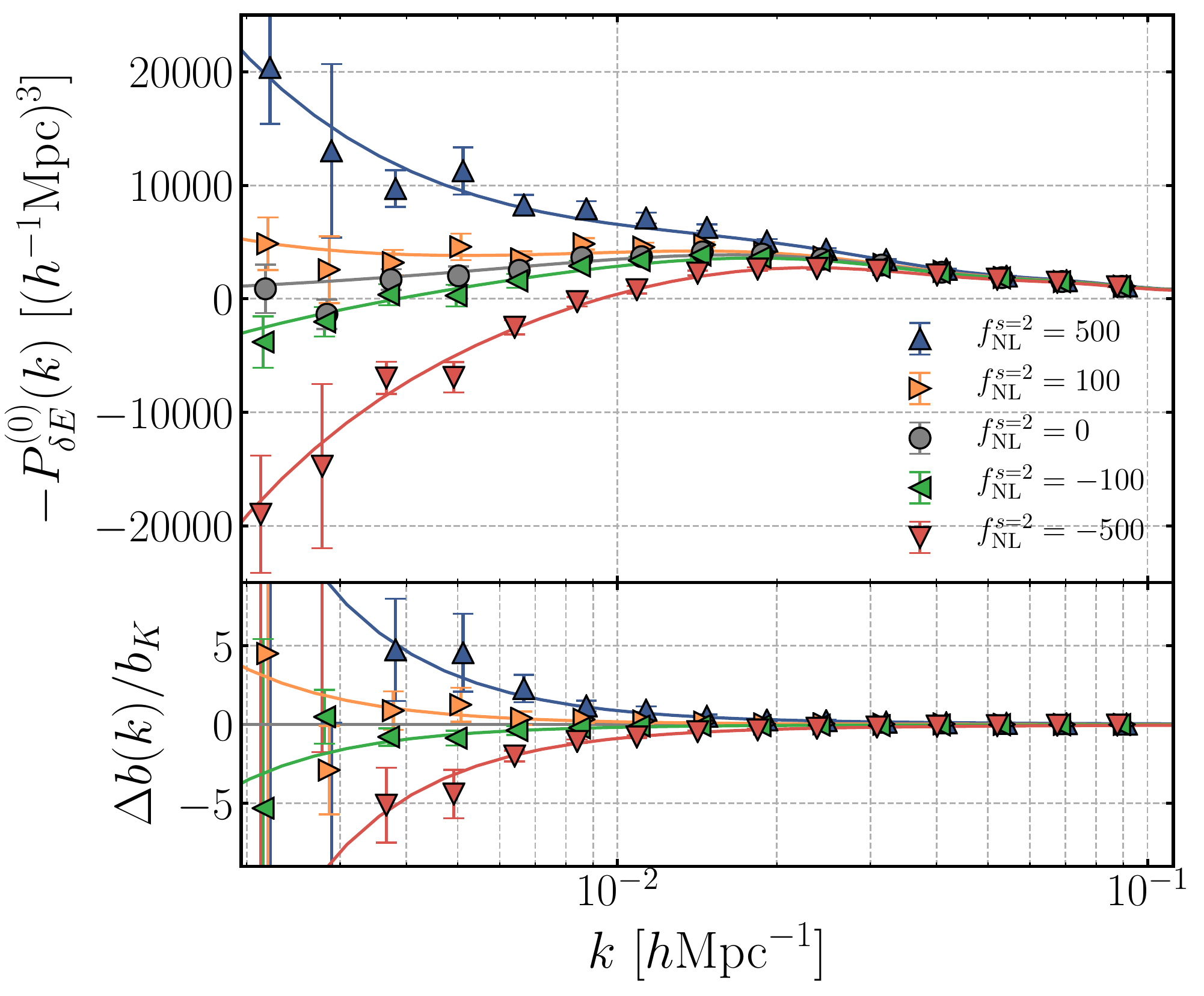}
        \caption{Similar to the middle panel of the previous figure, but the plot shows $P_{\delta E}^{(0)}(k)$ for the anisotropic PNG model 
        with $f_{\rm NL}^{s=2}=-500, -100, 100$ or $500$. For comparison, the gray points show the result for the Gaussian initial condition. 
        The solid lines show the best-fit model predictions (Eq.~\ref{eq:gammaij_PNG_model}).
        }
        \label{fig:fnls}
    \end{center}
\end{figure}
\begin{figure}
    \begin{center}
        \includegraphics[width=1.0\columnwidth]{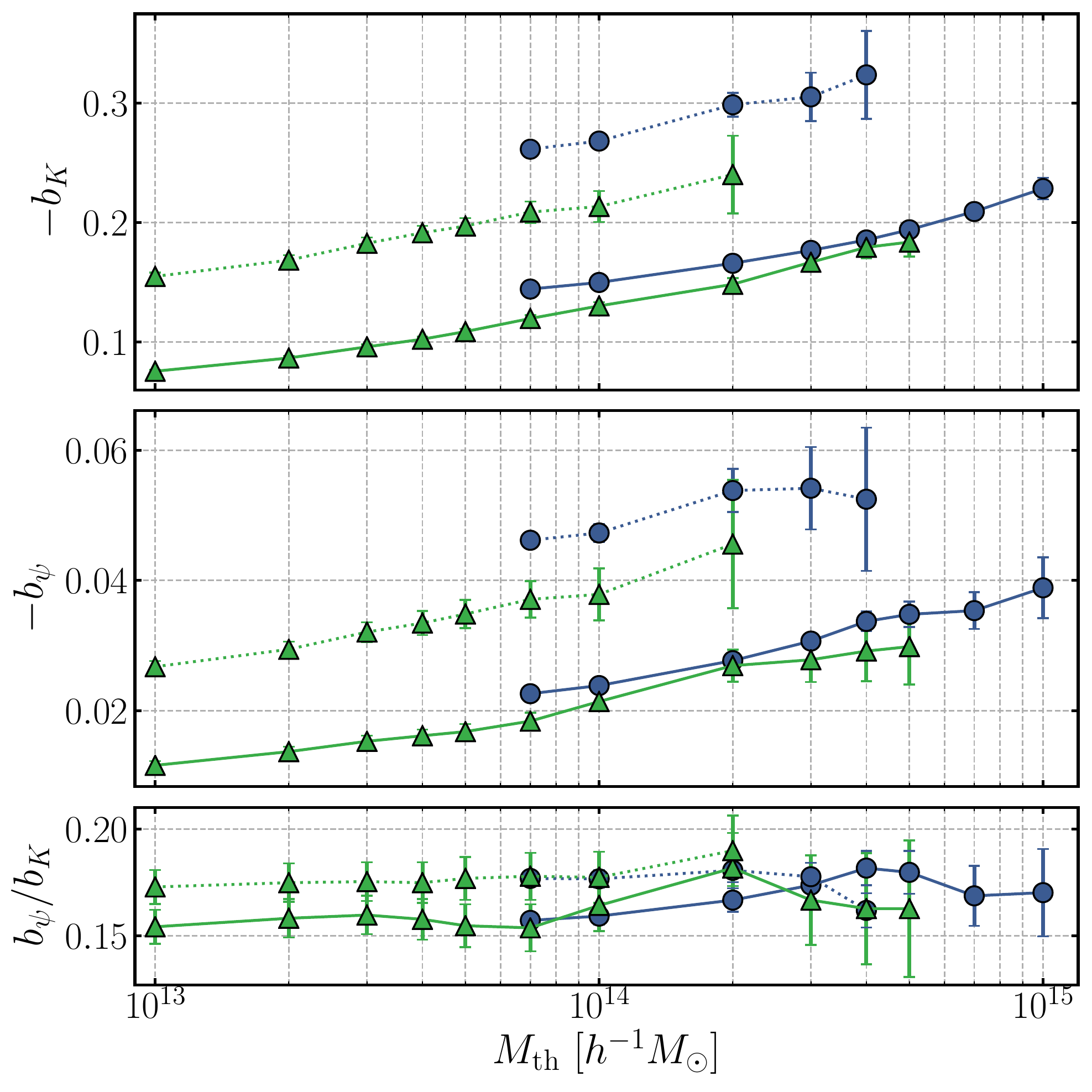}
        \caption{
        The best-fit IA parameters $b_K$ (top panel), $b_\psi$ (middle) and the ratio $b_\psi/b_K$ (bottom) for different mass-threshold samples of halos, selected with $M_{\rm vir} > M_{\rm th}$, at redshifts $z=0$ (solid line) and $1.0$ (dotted), respectively.
        \HL{
        The blue (circle) points and the green (triangle) points are obtained from $4.096~h^{-1}{\rm Gpc}$ and $2.048~h^{-1}{\rm Gpc}$ simulations, respectively.
        }}
        \label{fig:b_phi}
    \end{center}
\end{figure}
\begin{figure*}
    \begin{center}
        \includegraphics[width=1.9\columnwidth]{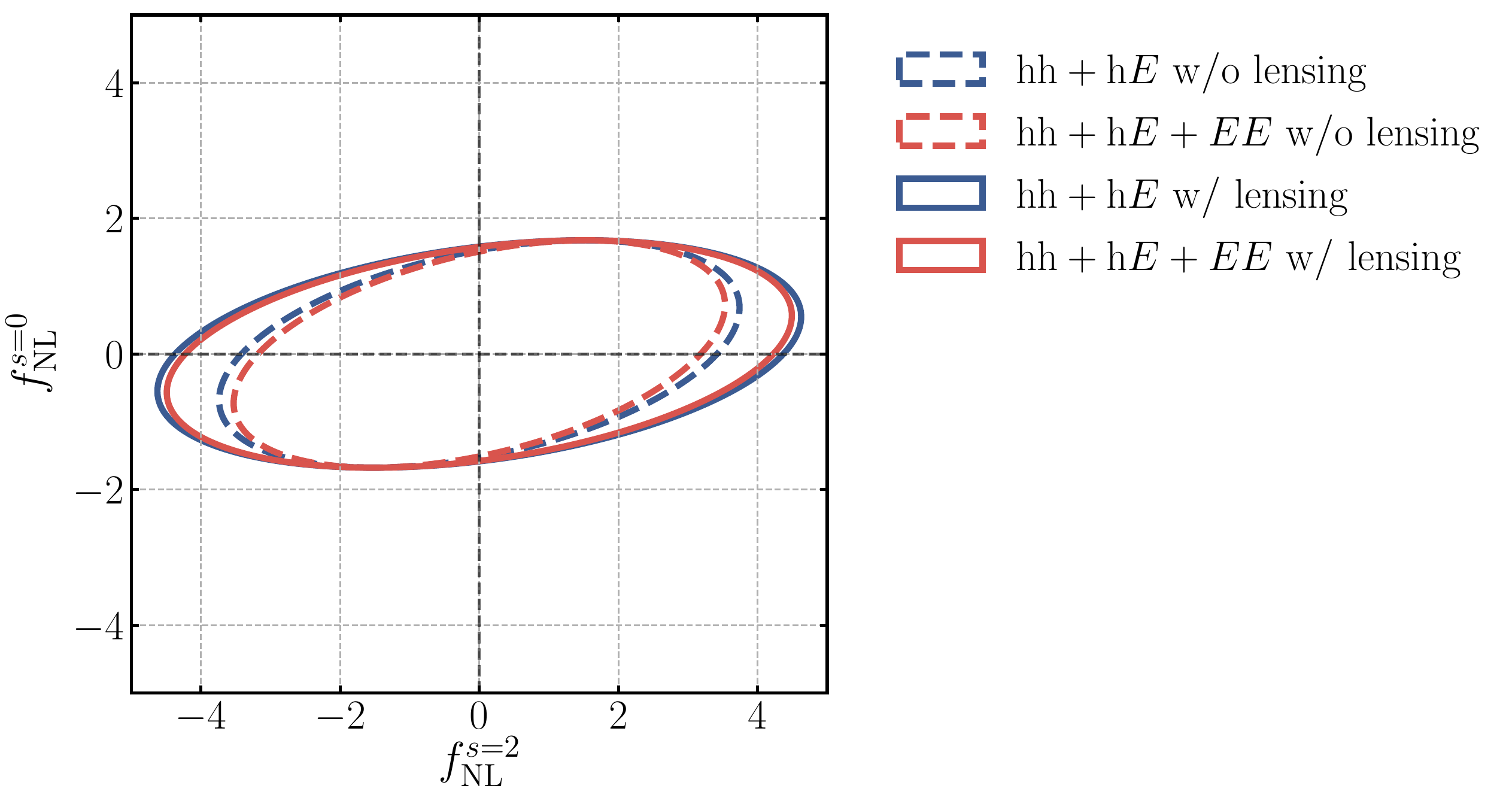}
        \caption{\HL{The 
        1$\sigma$ error contours in a two-parameter space of the anisotropic and isotropic 
        PNG parameters $(f_{\rm NL}^{s=2},f_{\rm NL}^{s=0})$, expected for a hypothetical survey volume of $69~(h^{-1}{\rm Gpc})^3$ corresponding to a galaxy survey of the redshift range $z=[0.5,1.4]$ and the solid-angle coverage $f_{\rm sky}=0.7$. Here we consider, as a proxy of galaxy sample, a mass-threshold sample of halos with $M_{\rm vir}\ge 10^{13}~h^{-1}M_\odot$, which have the number density of $2.9\times 
        10^{-4}~(h^{-1}{\rm Mpc})^{-3}$ that is comparable with that for luminous red galaxies.
        Here we employ $f_{\rm NL}^{s=0}=f_{\rm NL}^{s=2}=0$ for the fiducial values, i.e. the Gaussian initial conditions. Therefore the 
        size of contours displays the precision of such a galaxy survey for discriminating the PNG initial conditions; if the universe has the PNG condition outside the contours, the galaxy survey can detect the PNG condition  at more than $1\sigma$ from the measured IA and density 
        power spectra. Here we show the results for four cases: the blue- and red-solid contours 
        show the results obtained using  
        $P_{\rm hh}$ and $P_{{\rm h}E}$ with and 
        without the lensing contribution to the covariance matrix (see text for details). 
        The respective dashed contours show the results further including $P_{EE}$. 
        combining $P_{\rm hh}$, $P_{{\rm h}E}$, and $P_{EE}$ without lensing (dashed red) and with the lensing (solid red). 
        For the Fisher matrix calculation we employ 
        $k_{\rm min} = 0.002$ and $k_{\rm max}=0.1~h{\rm Mpc}^{-1}$ for the minimum and 
        maximum wavenumbers, respectively. }
        }
        \label{fig:fisher}
    \end{center}
\end{figure*}

\subsection{IA measurements}
To measure the IA correlations from simulations, we use a novel method developed in Ref.~\cite{Kurita_etal:2020}.
First we measure the inertia tensor defined by member particles of each halo according to 
$I_{ij}=\sum_{p} \Delta x_p^i \Delta x_p^j$, where $\Delta x_p^i$ is the relative position of each member particle from the halo center.
Taking the $z$-axis
to a hypothetical
line-of-sight direction, we define the two ellipticity components, $\epsilon^h_1,\epsilon^h_2$, for each halo from the ($2\times 2$) sub-matrix of $I_{ij}$ in the $xy$-plane as  
an observable halo shape
on the sky. 
After that, we use the Nearest-Grid-Point assignment \cite{Hockney_Eastwood:book} to define
the 
3-dimensional ellipticity fields, $\epsilon_1(\bx),\epsilon_2(\bx)$, as well as the matter and halo density fluctuation
fields, $\delta(\bx)$ and $\delta_{\rm h}(\bx)$.
Since the two ellipticity components 
form spin-2 fields in the $xy$-plane, we can perform the $E/B$-mode decomposition in Fourier space:
$E(\bk) \equiv \epsilon_1(\bk)\cos2\varphi_k + \epsilon_2(\bk) \sin2\varphi_k$ and
$B(\bk) \equiv -\epsilon_1(\bk)\sin 2\varphi_k + \epsilon_2(\bk) \cos 2\varphi_k$,
where $\varphi_k$ is the azimuthal angle of $\bk$. 

We then
estimate the IA power spectra from each simulation, and in this paper we mainly focus on the IA cross-power spectrum, defined as
\begin{align}
    \avrg{\delta(\bk) E(\bk')} \equiv (2\pi)^3\delta^3_{\rm D}(\bk+\bk')P_{\delta E}(k, \mu),
\end{align}
where $\mu \equiv \hat{\bk} \cdot \hat{\bz}$.
The linear alignment model with the anisotropic PNG predicts 
that the cross-power spectrum at small $k$ is given by
\begin{align}
P_{\delta E}(k,\mu)=(b_K+\Delta b_K) (1-\mu^2)P_\delta(k)
\label{eq:P_dE_PNG}
\end{align}
where $\Delta b_K=12b_\psi f_{\rm NL}^{s=2}{\cal M}^{-1}(k)$ from Eq.~(\ref{eq:gammaij_PNG_model})
and $P_\delta(k)$ is the matter auto-power spectrum.
Similarly the auto-power spectrum of IA $E$-field is given as
$P_{EE}(k,\mu)=(b_K+\Delta b_K)^2(1-\mu^2)^2P_\delta(k)$ at small $k$.
\HL{
The correlators including $B$-field vanish at linear order: 
$P_{\delta B}=P_{BB}=0$.}
The $\ell$-th multipole 
moment of the power spectrum at wavenumber $k$ is
defined as
\begin{equation}
P^{(\ell)}_{\delta E}(k)=\frac{2\ell+1}{2}\int_{-1}^{1}\!\mathrm{d}\mu~ {\cal L}_\ell(\mu)P_{\delta E}(k,\mu).
\end{equation}
In practice we use a discrete summation, instead of the integral, %
over the grid points in Fourier space corresponding to
each $k$-bin, spaced by the fundamental mode
$k_f=2\pi/L$ ($L$ is the side length of a simulation box) for the measurement of the spectra from simulation realizations. 
We similarly estimate, from each simulation, the halo-matter power spectrum, $P_{\delta{\rm h}}$, and 
the multipole 
moments for the auto-spectrum of the halo shape $E$-field, $P_{EE}$, and for
the cross-power spectrum between the $E$-field and 
the halo number density field, $P_{{\rm h}E}$.  
We set the minimum wavenumber, $k_{\rm min} = 0.002~h{\rm Mpc}^{-1}$, and adopt
the bin
width; $\Delta {\rm ln}k = 0.26$ (10 bins in one decade of $k$).
In this paper we do not include the redshift-space distortion effect due to peculiar velocities of halos for simplicity.

\section{Results}
\label{sec:results}

The middle and right panels of Fig.~\ref{fig:ps} show the main result of this paper.
The PNG simulation confirms that 
the anisotropic ($s=2$) PNG induces a scale-dependent modification 
in the IA power spectra in small $k$ bins in the linear regime compared to the Gaussian simulation, 
but does not change the halo-matter power spectrum, $P_{\delta{\rm h}}$ around the same scale shown in the left panel. 
On the other hand, the isotropic ($s=0$) PNG does not alter the IA power spectra, but does alter $P_{\delta{\rm h}}$ as
shown in Ref.~\cite{Dalal_etal:2008}. 
Thus, the scale-dependent bias of the IA power spectra 
is a unique feature originating from the anisotropy in the PNG,
hence, if detected, would serve as a smoking gun evidence
of the $s=2$ PNG.
\HL{
For all cases we confirmed that after the zero lag subtraction the $B$-mode auto- and cross-power spectra are consistent with zero within errors on large scales, which means that all the $B$-mode power spectra are not affected by both PNGs.}

In Fig.~\ref{fig:fnls} we compare the best-fit model predictions with the simulated IA power spectra 
for different values of $f_{\rm NL}^{s=2}$. To estimate the best-fit model, we first estimate $b_K$ in Eq.~(\ref{eq:gammaij_Gaussian}) by 
comparing $P_{\delta E}$ and $P_\delta$ up to $k=0.05$~\hMpci for the Gaussian simulation 
assuming the Gaussian covariance.
Then we estimate $b_\psi$ in Eq.~(\ref{eq:gammaij_PNG_model}) 
in the same way by using the simulated spectra
measured from all the PNG simulations with different $f^{s=2}_{\rm NL}$ values up to $k=0.05$~\hMpci, 
varying $b_\psi$  
as
the only free parameter. 
The figure shows that the best-fit model predictions 
are consistent with
the data points within the error bars. 

Fig.~\ref{fig:b_phi} shows the estimated $b_K$ and $b_\psi$ for different mass-threshold samples of halos at different redshifts. 
\HL{The results for different box-size simulations are not in perfect agreement with each other. 
This would be ascribed to the dependence of halo shape estimation, $I_{ij}$, on the number of member particles even for halos of a fixed mass scale, 
as discussed in Appendix~C of Ref.~\cite{Kurita_etal:2020} and Ref.~\cite{Akitsu:2020fpg}.
Nevertheless we find that the ratio of $b_\psi/b_K$ is not sensitive to the simulation resolution. 
The ratio does not vary with halo samples and redshifts significantly, displaying 
$b_\psi/b_K\sim 0.17$  for all the 
cases shown in the plot.
Hence we believe that the following results obtained assuming 
a ratio around this value
would be
robust against 
the numerical resolution issue.
The same sign of $b_K$ and $b_\psi$ implies that 
the response of halo shapes to the large-scale tidal field ($b_K$) is similar to that to the quadrupolar modulation in the small-scale fluctuations ($b_\psi$); an initial density peak is likely to collapse first in the direction of the largest eigenvector of $K_{ij}$ and $\psi_{ij}$.}

Now we estimate the ability
of a wide-area galaxy survey to constrain
the anisotropic PNG amplitude, using the Fisher information matrix:
\begin{align}
F_{\alpha\beta}
=\sum_{XX'}\sum_{\ell \ell'}\sum_{k_i,k_j}
\frac{\partial P^{(\ell)}_{X}(k_i)}{\partial \alpha}
\left[{\bf C}\right]^{-1}_{XX'(\ell\ell')ij}
\frac{\partial P^{(\ell')}_{X'}(k_j)}{\partial\beta},
\label{eq:fisher}
\end{align}
where $(X,X')=\{{\rm hh}, {\rm h}E, EE\}$, 
$(\alpha,\beta) = \{f_{\rm NL}^{s=0}, f_{\rm NL}^{s=2}\}$, and ${\bf C}$ is the covariance matrix between $P^{(\ell)}_{X}(k_i)$ and $P^{(\ell')}_{X'}(k_j)$ for which we assume a Gaussian covariance 
taking into account the shot noise and the intrinsic shape noise measured from the simulations
\cite{Kurita_etal:2020}. 
\HL{Note that, in an actual observation, the auto-power spectrum of galaxy shapes, $P_{EE}(k)$, 
receives a contribution from the weak lensing distortion effects on the shapes due to foreground large-scale structures at different redshifts along the line-of-sight direction -- {\it cosmic shear}. 
Since the foreground structures are safely considered to be uncorrelated with the large-scale structure causing the IA effect, cosmic shear only adds statistical errors to the measurement of $P_{EE}(k)$. 
In Appendix~\ref{app:WL},
we describe
how to calculate the cosmic shear
contribution to $P_{EE}(k)$.}

To perform the Fisher forecast, 
we consider a hypothetical survey covering a comoving volume of $V_s=69~(h^{-1}{\rm Gpc})^3$ that roughly corresponds to a spectroscopic survey 
with sky coverage $f_{\rm sky}\simeq 0.7$ 
in the redshift range $z=[0.5,1.4]$.
We then consider a mass-threshold halo sample with $M_{\rm vir}\ge 10^{13}~h^{-1}M_\odot$ at $z=1$. 
From the simulation we find that such a halo sample has 
$2.9\times 10^{-4}~(h^{-1}{\rm Mpc})^{-3}$ for the number density. 
This sample roughly corresponds to a sample of luminous early-type galaxies that reside in massive 
halos of $\sim 10^{13}h^{-1}M_\odot$ \citep{2015ApJ...806....2M}.  
For the Fisher forecast we need to adopt the fiducial values for the model parameters. 
We first employ, as the fiducial cosmology, the cosmological parameters that are consistent 
with the {\it Planck} cosmology \citep{Planck2015_cosmo}.
For the PNG parameters (Eq.~\ref{eq:bispectrum_s}), we employ $f^{s=2}_{\rm NL}=f^{s=0}_{\rm NL}=0$ as the fiducial values, i.e. 
the Gaussian initial condition. Nevertheless we need to model the density and IA power spectra 
of halos to compute variations in the power spectra for finite changes in each of the model parameters
($f_{\rm NL}^{s=2},f_{\rm NL}^{s=0}$) from the zero value. 
To model the IA effect for $f_{\rm NL}^{s=2}$, we use Eq.~(\ref{eq:gammaij_PNG_model}), where 
we employ $b_K=-0.15$ and $b_\psi=-0.027$ for the fiducial values of bias parameters, 
as inferred from Fig.~\ref{fig:b_phi}.
The assumed value of $b_K$ is roughly consistent with that found from a sample of 
massive, early-type galaxies with the similar comoving number density to $10^{-4}~(h^{-1}{\rm Mpc})^3$  
in the Illustris-TNG simulation \cite{2020arXiv200900276S}.
\HL{For the halo auto-power spectrum 
in the models with isotropic PNG 
, 
we employ 
$P_{\rm hh}(k) = (b_1 + \Delta b_1)^2P_\delta(k)$
with $\Delta b_1 = 2(b_1-1)\delta_{\rm cr} f_{\rm NL}^{s=0}{\cal M}^{-1}(k)$, where 
$\delta_{\rm cr}=1.686$
and we use $b_1 = 2.9$ for the linear density bias parameter that is measured from the Gaussian 
simulation.}
For the maximum wavenumber in the Fisher matrix calculation (Eq.~\ref{eq:fisher}), we throughout this paper adopt 
$k_{\rm max}=0.1~h{\rm Mpc}^{-1}$.

\HL{Fig.~\ref{fig:fisher} shows the
68\% errors for the PNG parameters expected for the survey setting described above.
Here we consider four cases in total; the results 
using either the combination of two
power spectra, $\{P_{\rm hh}, P_{{\rm h}E}\}$ or all the three spectra, $\{P_{\rm hh}, P_{{\rm h}E}, P_{EE}\}$, 
with or without the weak lensing contribution to the covariance matrix.
First, as is clear from Fig.~\ref{fig:fisher}, 
the two parameters ($f^{s=0}_{\rm NL}$ and $f^{s=2}_{\rm NL}$)  
can be simultaneously constrained from the combined measurements of $P_{\rm hh}$ and $P_{{\rm h}E}$.
Although the effects of the $s=0$ and $s=2$ PNGs to the scale-dependent bias in $P_{{\rm h}E}$ are degenerate with each other, the degeneracy can be broken by adding $P_{\rm hh}$, which is solely dependent on the $s=0$ PNG.
Second, 
the $E$-mode auto-power spectrum plays a little role
in determining PNGs.
This is consistent with the fact that the signal-to-noise ratio for the $E$-mode auto-power spectrum is much smaller than that of $P_{{\rm h} E}$ 
\citep[also see][for the similar discussion]{Kurita_etal:2020}.
Third, the errors of $f_{\rm NL}^{s=2}$
are only slightly degraded when
taking into account the lensing contribution in the covariance matrix.}
To summarize, a wide-area galaxy survey enables us to
obtain the precision 
$\sigma(f_{\rm NL}^{s=2})\simeq 4$ or $\sigma(b_{\psi}f^{s=2}_{\rm NL})\simeq 0.1$.
Note that, if we change the minimum wavenumber to $k_{\rm min}=0.005~h{\rm Mpc}^{-1}$ from our default choice of 
$k_{\rm min}=0.002~h{\rm Mpc}^{-1}$, the precision 
is slightly degraded to $\sigma(f_{\rm NL}^{s=2})\simeq 5$.
These results suggest that the anisotropic PNG can be detected at more than $1\sigma$, if the 
true value 
of $f_{\rm NL}^{s=2}$ is larger than $\sim 5$
by a wide-area galaxy survey with a setting similar to that considered here.
The precision of the IA power spectrum
is much better than the forecast in Ref.~\cite{Schmidt_etal:2015} which is based on the angular IA power spectrum instead of the 3D 
IA power spectrum. \HL{This improvement reflects the power of the 3D power spectrum, which
allows us to access
much more
Fourier modes than in the 2D angular power spectrum.}
Furthermore, 
this 
result is better than the current CMB constraint, $\sigma(f_{\rm NL}^{s=2})\simeq 19$ \cite{Planck_PNG:2019}.
\HL{We also note that the constraint is degraded to $\sigma(f_{\rm NL}^{s=2})\sim 40$, still comparable to the current CMB constraint, even if we assume tracers with a weaker response to the large-scale tidal field and the anisotropic PNG by a factor of ten,
$b_K=-0.015$, instead of $b_K=-0.15$.}
In any case it should be noted that the IA method constrains the anisotropic PNG at different redshifts and for different length scales 
compared to the CMB constraints, and the two methods are 
complementary to each other.

\section{Discussion}
\label{sec:discussion}

In this paper 
we have shown that the IA power spectra, measured from the wide-area spectroscopic and imaging surveys of galaxies for the same region of the sky, can be used to constrain the anisotropic PNG at a precision comparable 
\tnrv{to}
or even better than 
the current CMB constraint. Here an imaging survey is needed to measure shapes of individual galaxies, while a spectroscopic survey is needed to obtain
their three-dimensional positions.
A further improvement can be obtained, e.g. by having a larger volume covering up to a higher redshift, 
combining the bispectrum information of both the number density \cite{Assassi_etal:2015} and IA, combining the IA power spectra of different density and shape tracers  (i.e. multi-tracer 
technique in Refs.~\cite{Seljak:2009,Chisari_etal:2016}), and also including the redshift-space distortion effect.
In addition, it is important to investigate effects of the anisotropic PNGs with a particular scale dependence \citep{Kogai_etal:2018}.
These are all interesting, and worth 
exploring in more detail.

\begin{acknowledgments}

We thank Maresuke~Shiraishi and Jingjing~Shi for useful discussions.
This work was supported in part by World Premier International
Research Center Initiative (WPI Initiative), MEXT, Japan, and JSPS
KAKENHI Grant Numbers 
JP15H05887, JP15H05893, JP15K21733, JP17K14273, JP19H00677, JP19J12254
JP20J22055, JP20H05850, and JP20H05855
and by JST AIP Acceleration Research Grant Number JP20317829, Japan.
K.A. and T.K. are supported by JSPS
Research Fellowship for Young Scientists.
Numerical computation was carried out on Cray XC50 at Center for Computational Astrophysics, National Astronomical Observatory of Japan.
\end{acknowledgments}


\appendix
\section{\HL{Cosmic shear contribution to the auto-power spectrum of galaxy shapes}}
\label{app:WL}

In this appendix we estimate the lensing contribution to the auto-power spectrum 
of galaxy shapes $P_{EE}(k)$, following Ref.~\cite{Hui_etal:2008}.
The observed $E$-mode field of galaxy at a 
given redshift 
is generally expressed by the
sum of the IA effect due to large-scale structure 
at the same redshift and the weak lensing effect due to {\it foreground} large-scale structures at different redshifts:
\begin{equation}
    E^{\rm obs}(\bx) = E^{\rm IA}(\bx) + E^{\rm lens}(\bx).
\end{equation}
Here $E^{\rm lens}(\bx)$ 
is the lensing convergence 
field at the position $\bm{x}$, denoted as 
$\kappa(\bx)$ in the usual notation \citep[e.g.][]{2001PhR...340..291B}, and is given by
the weighted line-of-sight integration of the matter density fluctuation field along the line-of-sight
direction:
\begin{align}
    \kappa(\bx) &= \frac{3}{2}H_0^2\Omega_{\rm m0} \int_0^\chi \dd \chi' \frac{(\chi - \chi')\chi'}{\chi}(1+z') \delta(\chi', \boldsymbol{\theta})\\
    &\equiv \int_0^\chi \dd\chi' \mathcal{K}(\chi,\chi') \delta(\chi', \boldsymbol{\theta}),
\end{align}
where $\delta(\chi, {\bm \theta})=\delta(\bx)$ is the matter density fluctuation field and 
$\mathcal{K}(\chi, \chi')$ is the lensing kernel function.
Note that $\chi$ is the comoving distance to galaxies that are used for the IA measurement, and 
$\chi'$ denotes the distance to the lensing matter distribution satisfying 
$\chi'<\chi$. The lensing effect builds up over Gpc scales that are much longer than the correlation 
length of the IA power spectrum. Hence the lensing effect acts as a ``statistical'' noise to 
the IA measurement because of $\langle E^{\rm IA} E^{\rm lens}\rangle=0$ (or 
$\langle E^{\rm IA}\kappa\rangle=0$).
The auto-correlation function of $\kappa$ can be computed as
\begin{align}
    &\xi_{\kappa \kappa}(\bx_1,\bx_2) \nonumber\\ 
    &= 
    \int_0^{\chi_1} \dd\chi'_1 \mathcal{K}(\chi_1,\chi'_1) 
    \int_0^{\chi_2} \dd\chi'_2 \mathcal{K}(\chi_2,\chi'_2) \nonumber\\
    &\hspace{1cm} \times
    \langle \delta(\chi'_1, \boldsymbol{\theta}_1) \delta(\chi'_2, \boldsymbol{\theta}_2) \rangle,\\
    &\simeq 
    \int_0^{{\rm min}(\chi_1,\chi_2)} \dd\chi' \mathcal{K}(\chi_1,\chi')\mathcal{K}(\chi_2,\chi') \nonumber\\
    &\hspace{1cm} \times \int \frac{\dd^2 \bk_\perp}{(2\pi)^2} P_\delta (k_\perp ; z') e^{i \bk_\perp \cdot \chi' (\boldsymbol{\theta}_1 -\boldsymbol{\theta}_2)}, \nonumber\\
    &\simeq 
    \int_0^{\bar{\chi}} \dd\chi' \mathcal{K}^2(\bar{\chi},\chi') \nonumber\\
    &\hspace{1cm} \times \int \frac{\dd^2 \bk_\perp}{(2\pi)^2} P_\delta (k_\perp ; z') e^{i \bk_\perp \cdot \chi' (\boldsymbol{\theta}_1 -\boldsymbol{\theta}_2)},
    \label{eq:xi_kappa}
\end{align}
where $\bk_\perp$ denotes 
the component of the wavevector perpendicular to the line-of-sight direction
and 
we have used the Limber approximation \citep{1953ApJ...117..134L}
in the second line. 
In the third line on the r.h.s. we set $\chi_1 \simeq \chi_2 \equiv \bar{\chi}$ assuming that 
galaxies used for the IA measurement are much more distant than typical lens redshifts:
$\chi_1,\chi_2\gg \chi_1^\prime,\chi_2^\prime$. This is a good approximation if we consider a high redshift for IA galaxies as we have assumed in this paper. Hence
we approximate $\chi_1\simeq \chi_2\simeq \bar{\chi}$, where $\bar{\chi}$ is the mean distance to IA galaxies.
The Fourier transformation of Eq.~(\ref{eq:xi_kappa}) yields
\begin{equation}
    P_{\kappa \kappa}(\bk) = |W(k_\parallel)|^2 \bar{\chi}^2 C^{\kappa \kappa}_{\ell = k_\perp \bar{\chi}}, 
    \label{eq:Pwl}
\end{equation}
where $W$ is a window function 
in the radial direction that characterizes a radial selection of IA galaxies
(we ignore a window function of 
transverse components for simplicity). The angular power spectrum of the lensing field,
$C^{\kappa\kappa}_\ell$, is given by
\begin{equation}
    C^{\kappa \kappa}_\ell = \int_0^{\bar{\chi}} d\chi' \frac{\mathcal{K}^2(\bar{\chi},\chi')}{\chi'^2} P_\delta (\ell/\chi'; z').
\end{equation}
To evaluate $W(x_\parallel)$, we assume a top-hat function as
\begin{equation}
    W(x_\parallel) = 1/\sqrt{L}, \text{~~~~~~if~$\bar{\chi}-L/2 < x < \bar{\chi}+L/2$},
\end{equation}
and otherwise $W(x_\parallel)=0$.
Using Eq.~(\ref{eq:Pwl}), we can calculate the weak lensing contribution to 
the Gaussian term in the covariance matrix of IA auto-power spectrum, $P_{EE}$.
Note that we confirmed that our results of the Fisher analysis in Fig.~\ref{fig:fisher} are almost independent to the choice of the window length in the range of $L=[300,1000]~h^{-1}{\rm Mpc}$ in the case that a typical redshift of IA galaxies is $z\sim 1$ as we have assumed in the main text.

\bibliography{refs}

\end{document}